\documentclass[12pt]{article}
\pdfoutput=1 
\usepackage{graphicx,amsmath}
\usepackage{units}
\usepackage{color}
\usepackage{float}

\newlength{\dinwidth}
\newlength{\dinmargin}
\def\be{\begin{equation}}    
\def\ee{\end{equation}}  
\def\bea{\begin{eqnarray}}                      
\def\eea{\end{eqnarray}}
\def\ch1{$\chi(1^+)$}
\def\lapproxeq{\lower .7ex\hbox{$\;\stackrel{\textstyle                                                    
<}{\sim}\;$}}                                                    
\def\gapproxeq{\lower .7ex\hbox{$\;\stackrel{\textstyle                                                    
>}{\sim}\;$}}

\begin{document}

\begin{flushright}                                                    
IPPP/18/94  \\                                                    
\today \\                                                    
\end{flushright} 

\vspace*{0.5cm}

\begin{center}

{\Large \bf Absorptive effects and power corrections} \\
\vspace{0.5cm}
{\Large \bf in low $x$ DGLAP evolution}\\
\vspace{1.0cm}

M.R. Pelicer$^a$, E.G. de Oliveira$^{a,b}$, A.D. Martin$^b$ and  M.G. Ryskin$^{b,c}$\\ 

\vspace{0.5cm}
$^a$ {\it Departamento de F\'{i}sica, CFM, Universidade Federal de Santa
Catarina, C.P. 476, CEP 88.040-900, Florian\'opolis, SC, Brazil}\\    
$^b$ {\it Institute for Particle Physics Phenomenology, University of Durham, Durham, DH1 3LE } \\
$^c$ {\it Petersburg Nuclear Physics Institute, NRC Kurchatov Institute, Gatchina, St.~Petersburg, 188300, Russia } \\ 

\begin{abstract}

We address a long standing problem concerning the scale behaviour of parton densities in the low $x$, low $Q^2$ domain. We emphasize the important role of absorptive corrections at low $x$ and use knowledge of diffractive deep inelastic scattering to exclude the absorptive effect from conventional deep inelastic data. In this way we obtain a significantly different low $x$ behaviour of the gluon density, which is now much better described by linear DGLAP evolution. Accounting also for a second power correction, which arises from the freezing of $\alpha_s$ at low $Q^2$, leads to an essentially flat behaviour of the low $x$ gluon density.

\end{abstract}

\end{center}
\vspace{0.5cm}

\section{Introduction}

\begin{figure} [tbh]
	\begin{center}
		\vspace*{0.0cm}
		\includegraphics[width=14cm]{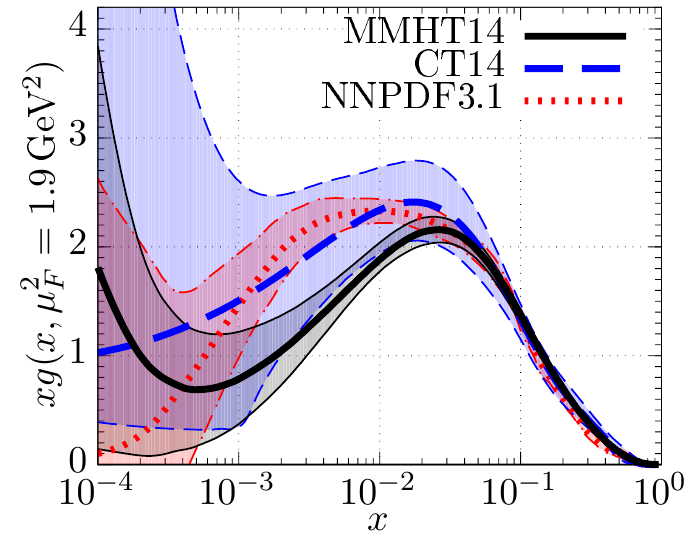}
		\vspace{-0.0cm}
		\caption{\sf Low $Q^2 = 1.9$\,GeV$^2$ gluon distributions obtained via the global parton analyses of MMHT14~\cite{MMHT14}, CT14~\cite{CT14}, and NNPDF3.1~\cite{NNPDF3.1} using LHAPDF~\cite{LHAPDF}. }
		\label{fig:pdf}
	\end{center}
\end{figure}

The conventional DGLAP evolution does not describe the deep inelastic scattering data in the low $x$, low $Q^2$ region very well. In fact the conventional PDF fits to the `global' data show that the gluon is not well determined in this domain, as illustrated in Fig.~\ref{fig:pdf}. Recent detailed studies of the combined H1 and ZEUS HERA deep inelastic data \cite{HERAdata} have been presented in \cite{MMHT,Foster}. They show that the description of the data can be improved by allowing for phenomenological  power corrections to the structure functions of the form
\be
F_i(x,Q^2)~\to ~ F_i(x,Q^2)(1+a_i/Q^2),
\ee
with $i=2,L$, where $a_L\simeq 4$ GeV$^2$ is the most important parameter. Note
that in such a parametrization the power correction does not
depend on $x$.
  
  From a more theoretical viewpoint it is known that in the low $x$, low $Q^2$ region absorptive corrections (or gluon recombination effects) are not negligible and reduce the growth of the gluon parton distribution function (PDF).  These effects were first emphasized long ago by Gribov-Levin-Ryskin \cite{GLR} and by Mueller-Qiu \cite{MQ} where an extra non-linear term, quadratic in the gluon density, was added to the linear DGLAP evolution equation for the gluon density
\be
\frac{\partial xg(x,Q^2)}{\partial {\rm ln}Q^2}~=~\frac{\alpha_s}{2\pi}\sum_{a'=q,g}P_{ga'}\otimes a'~-~\frac{9\alpha^2_s(Q^2)}{2R^2Q^2}\int^1_x\frac{dx'}{x'}[x'g(x',Q^2)]^2,
\label{eq:1}
\ee
where $R\sim1$ fm is of the order of the proton radius. The equation accounts for all `fan' diagrams. That is, all possible $2\to 1$ ladder recombinations are resummed to leading order of the parameter $\alpha_s$ln$(1/x)$ln$(Q^2/Q^2_0)$. It leads to saturation of the gluon density at low $Q^2$ with decreasing $x$. 
Other early works on this topic can be found in \cite{Bartels,Collins}. Nowadays a more precise non-linear evolution equation has been developed by Balitsky-Kovchegov \cite{BK-B,BK-K} based on BFKL evolution.

To investigate the role of the absorptive effects on the behaviour of the gluon in the low $x$ region, we first correct the low $x$ Deep Inelastic Scattering (DIS) data using the known diffractive DIS (dDIS) PDFs \cite {MRW,MRW1} and the AGK cutting rules \cite{AGK}. The resulting modified DIS data should now be driven by linear DGLAP evolution. Indeed we find the quality of the NLO fit is much improved. Formally the absorptive  effect behaves as a $1/Q^2$ correction (see, for example, eq.~(\ref{eq:1})) which becomes important due to the large gluon density at low $x$.

Besides absorption in the low $Q^2$ region the confinement effect is expected to modify the running of the QCD coupling $\alpha_s(Q^2)$. Again, for large $Q^2$, this formally plays the role of a power correction, which nevertheless may be important in describing the low $Q^2$ data.

Since confinement excludes an interaction at large distances, larger than the finite size of hadrons, it is actually impossible to reach a low value of the factorization scale.  Moreover, there are phenomenological arguments, partly confirmed by lattice and by Schwinger-Dyson calculations \cite{Aguilar:2001zy, Simonov:2001dt, Stefanis:2009kv}, that the value of the QCD coupling becomes frozen at a scale $\mu_0^2\sim 0.5$ GeV$^2$, and/or the singularity of the gluon propagator does not occur at $m_g=0$ but corresponds to an effective mass $m_g^2 \sim 0.5$ GeV$^2$. Therefore it is reasonable to freeze the DGLAP evolution somewhere in this region.  The simplest way to do this is to replace the argument ln$Q^2$ of DGLAP evolution by ln$(Q^2+\mu_0^2)$, or just to freeze only the value of QCD coupling by using the 
$\alpha_s(Q^2+\mu^2_0)$ in the place of $\alpha_s(Q^2)$. The latter procedure was proposed in~\cite{Shir} and implemented 
recently, for example, in \cite{Ter}. The value of the shift is typically $\mu_0\sim 1$ GeV.

Both the absorptive and confinement effects are studied here.

\section{AGK relations} 
\begin{figure} [htb]
\begin{center}
\vspace*{0.0cm}
\includegraphics[trim=4cm 9cm 0cm 2cm,scale=0.6, clip]{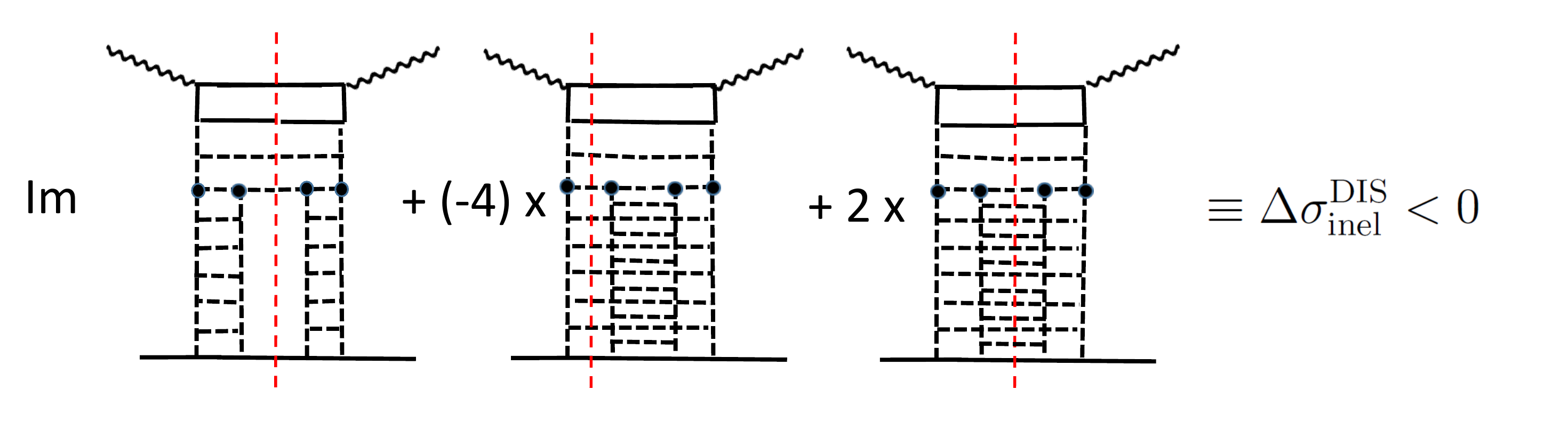}
\vspace{-1.0cm}
\caption{\sf The first absorptive correction due to the rescattering of an intermediate parton. The continuous and dashed lines correspond to quarks and gluons respectively. The sum of all possible cuts gives the imaginary part of the amplitude. According to the AGK cutting rules \cite{AGK} the weights of the three diagrams are $1,~-4,~2$ respectively. The negative weight of the central diagram, with only one ladder cut, follows since it describes the absorptive correction due to the rescattering of an intermediate parton.}
\label{fig:A}
\end{center}
\end{figure}

  \begin{figure} [htb]
\begin{center}
\vspace*{0.0cm}
\includegraphics[trim=7cm 8cm 5cm 1cm,scale=0.6, clip]{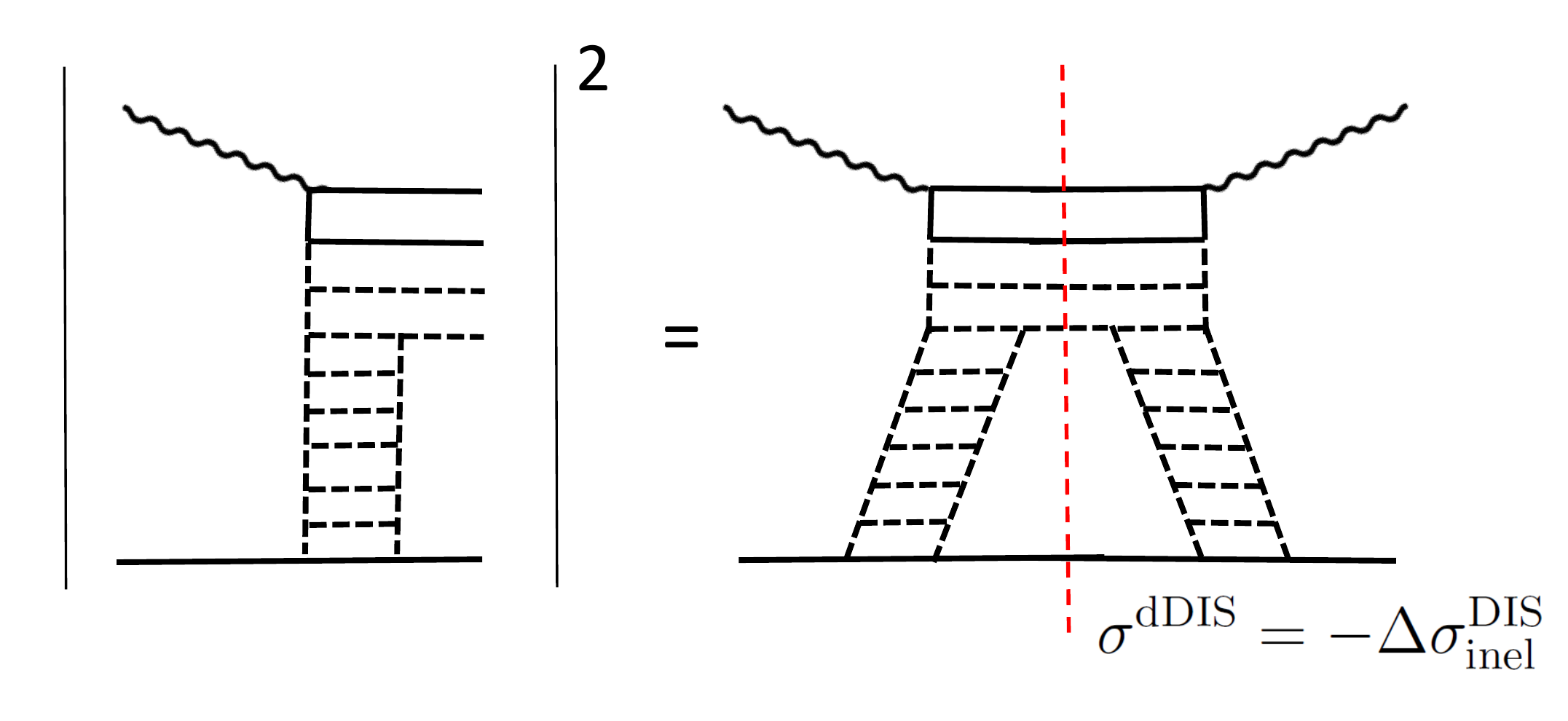}
\vspace{-0.0cm}
\caption{\sf A contribution to the cross section for diffractive DIS (dDIS). The left and right diagrams show $|A|^2$ and $AA^*$.}
\label{fig:B}
\end{center}
\end{figure}

 The simplest absorptive correction, $\Delta \sigma$, is shown in Fig.~\ref{fig:A}, where an intermediate parton has an additional interaction with the target proton. In another language, it is said that this diagram describes the fusion of different cascades into a single ladder which reduces the number of low $x$ partons.  The sum of all possible cuts of the diagram (depending on the position of the vertical line) give the imaginary part of the whole amplitude. The three diagrams shown contribute with different weights given by the AGK cutting rules \cite{AGK}. The weights are $1,~-4,~2$ respectively. The sum  
 \be
 +1-4+2=-1,
 \label{eq:AGK}
 \ee
 gives a total negative contribution to the cross section. An explicit calculation of the different cut contributions was given in \cite{BR,GLR}, while in \cite{AGK} these cutting rules were proved in general form.
 
On the other hand, diffractive DIS (dDIS) corresponds to the process where an incoming proton emits a Pomeron described by a DGLAP ladder. This ladder plays the role of the target for the observed DIS (as shown in the left diagram in Fig.~\ref{fig:B}). The corresponding cross section is shown in the diagram on the right. Note that the cut in Fig.~\ref{fig:B} corresponds to one of cuts in the triple-Pomeron diagram Fig.~\ref{fig:A}. According to the AGK cutting rules this cut has weight $+1$ in ({\ref{eq:AGK}).  That is, equal in value but opposite in sign to the sum of the cuts in Fig.~\ref{fig:A}. Therefore we can restore the DIS cross section $\sigma^{(0)}$ in the limit where the absorptive corrections are absent in the following way
 \be \label{eq:correction}
 \sigma^{(0)}~=~\sigma^{\rm DIS}_{\rm inel}~+~|\Delta \sigma |~=~\sigma^{\rm DIS}_{\rm inel}~+~\sigma^{\rm dDIS}.
 \ee
 
The resulting $\sigma^{(0)}$ should now satisfy linear DGLAP evolution\footnote{Strictly speaking, there are more complicated multi-Pomeron contributions with their own AGK relations. However, in the HERA region, already the first correction gives $\sigma^{\rm dDIS}/\sigma^{\rm full}\sim 10\% $, so the effect of more complicated diagrams may be neglected.}.

\section{Numerical implementation}   
To study the effects of absorption and of confinement in NLO PDF analyses at low $x$ (and low $Q^2$) we compare the PDF  fits of the original HERA data with those using modified data from which the absorptive effects are excluded, as explained above. We use the xFitter programme \cite{xFitter} with QCDnum~\cite{QCDNUM} evolution. We fit to the combined H1 and ZEUS inclusive DIS data \cite{HERAdata}, including both neutral and charged current,  together with canonical parametrizations for the gluon and quark densities at an input $Q^2=1.9$ GeV$^2$.  In particular the input gluon is parametrized as
\be
x g(x) = A x^B (1-x)^C - A' x^{B'} (1-x)^{25}.
\label{eq:p}
\ee
 The last term is contrived to allow the gluon to decrease at small $x$ and to have a negligible effect otherwise due to the large power of the $(1-x)$ factor \cite{MMHT14}. In practice we find the presence of this term is not favoured by the fits. Indeed, only in Fig.~\ref{fig:5} do we show fits which include the final term in (\ref{eq:p}).  The power $C' = 25$ is chosen so that the second term will only noticeably contribute to the gluon distribution for $x < 10^{-2}$ if $A \approx A'$. 
 
 As mentioned above, we explore the effect of absorption by modifying the HERA DIS data by subtracting the lowest absorptive contribution using the known MRW results for dDIS PDFs which had been obtained from their NLO analysis of the H1 diffractive DIS data for $Q^2>$8.5 GeV$^2$ \cite{H1diff}.  The resulting data should satisfy linear DGLAP evolution.  It is important to note that in this way we are able to allow for the inhomogeneous term in the DGLAP evolution of the diffractive PDFs; see (\ref{a1}) in the Appendix..  That is, the diffractive PDFs are not simply described by new input at some fixed $Q_0$ (taken, for example, from the H1 fit B to their diffractive DIS data) but the new contribution, described by the inhomogeneous term is added during the evolution in scale $\mu$. At each interval of $d\mu$ this contribution is proportional to $d\mu^2/\mu^4$: see the last term in (\ref{a1}) and the $1/\mu^2$ behaviour of the Pomeron flux $f_P$. Thus the inhomogeneous term produces a $1/\mu^2$ correction in the evolution in ln$\mu^2$.   
 
Recall that MRW fits only to the H1 diffractive data above 8.5 GeV$^2$ (as recommended by the H1 collaboration). However, since the expressions used in the MRW analysis correctly include the scale dependence of the inhomogeneous contribution, they should provide the proper extrapolation to a lower scale of about 2 GeV$^2$. Note also that the momentum fraction $z$ dependence was not fitted, but was explicitly given by perturbative QCD (see the Appendix for brief details of the MRW analysis of diffractive DIS data).
 
To explore the effect of confinement we repeated the analyses replacing the QCD coupling $\alpha_s(Q^2)$ by 
$\alpha_s(Q^2+\mu^2_0)$ for various values of $\mu_0^2$ \cite{Simonov:2001dt,Shir}.  The value of $\alpha_s(M_Z^2)$ was fixed to be 0.118.
Here we compare results for $\mu_0^2=0$, corresponding to no effects of confinement, with those obtained for $\mu_0^2=1$ GeV$^2$.

\section{Discussion of the results}

\begin{figure} [tbh]
	\begin{center}
		\vspace*{0.0cm}
		\includegraphics[width=14cm]{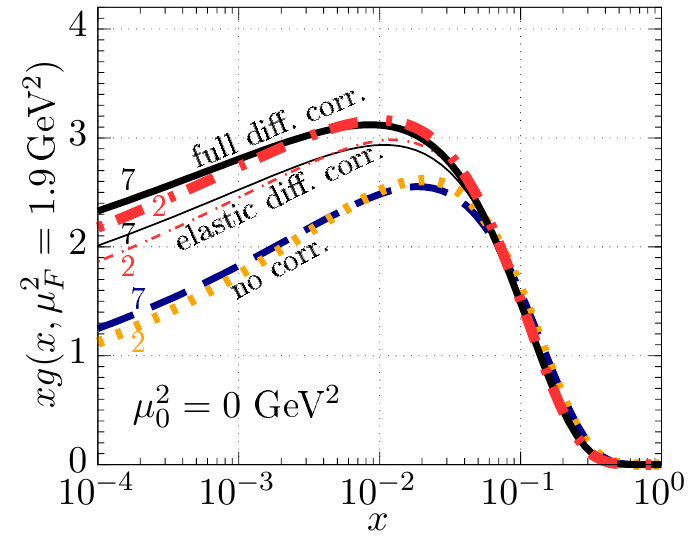}
		\vspace{-0.0cm}
		\caption{\sf 
			Gluon distributions at $\mu_F^2 = 1.9$\,GeV$^2$. We show three pairs of curves. In each pair the gluons differ according to whether the fit included DIS data with $Q^2>2$ or 7 GeV$^2$. The lower pair of curves (marked no correction) are the conventional gluons. If we account for the full diffraction corrections then we obtain at low $x$ noticeably larger gluons (the upper pair of curves). Finally the pair of thin curves correspond to the case where only the `elastic' dDIS data are used as a correction (that is data with the proton being left intact).}
		\label{fig:1}
	\end{center}
\end{figure}

We performed numerous fits to the HERA DIS data \cite{HERAdata}. In Figs.~\ref{fig:1}~$-$~\ref{fig:5} we show the minimum selection of the results that best describes the effects of including the power corrections, which arise from absorption and confinement, on the low $x$ behaviour of the gluon density. The quality of the fits is summarized in Table~\ref{tab:1}. The study concludes that the upper curve in Fig. \ref{fig:2} is the favoured behaviour of the low $x$ gluon PDF.

\begin{figure} [tbh]
	\begin{center}
		\vspace*{0.0cm}
		\includegraphics[width=8cm]{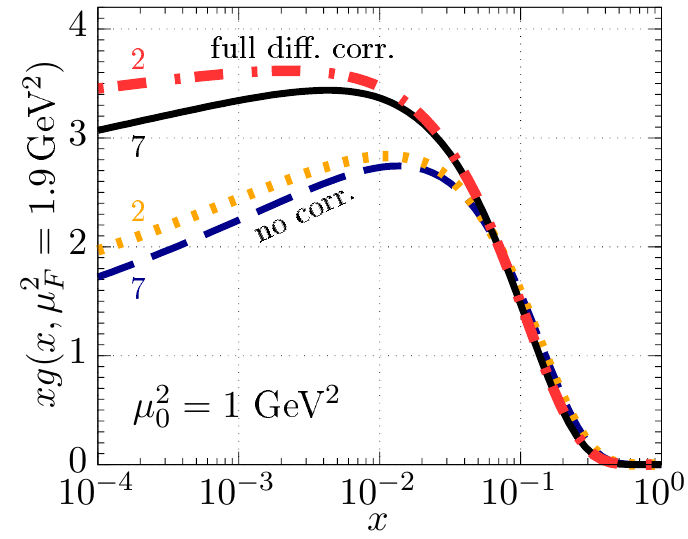}
		\includegraphics[width=8cm]{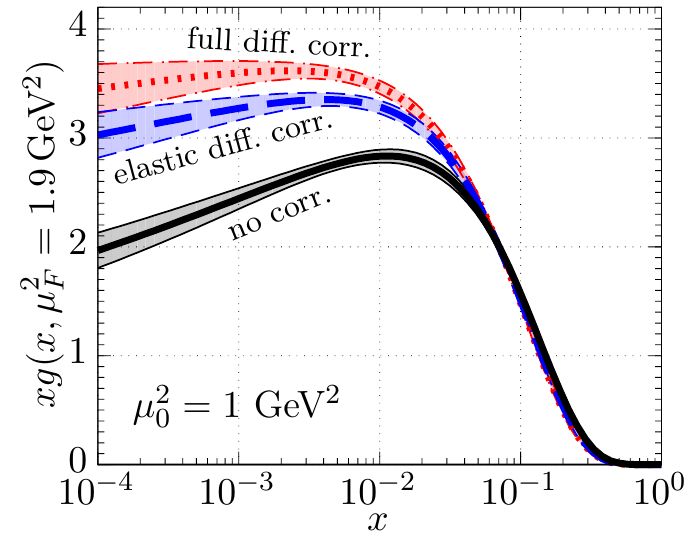}
		\vspace{-0.0cm}
		\caption{\sf Left: Gluon distributions fitted including power corrections coming from the low scale behaviour of running $\alpha_s$ coupling ($\mu_0=1$ GeV). The `linear' gluons become almost flat in the small $x$ region. Right: Gluon distributions with error bars using data with $Q^2 > 2$\, GeV$^2$ and $\mu_0=1$\,GeV.}
		\label{fig:2}
	\end{center}
\end{figure}

\begin{figure} [tbh]
	\begin{center}
		\vspace*{0.0cm}
		\includegraphics[width=14cm]{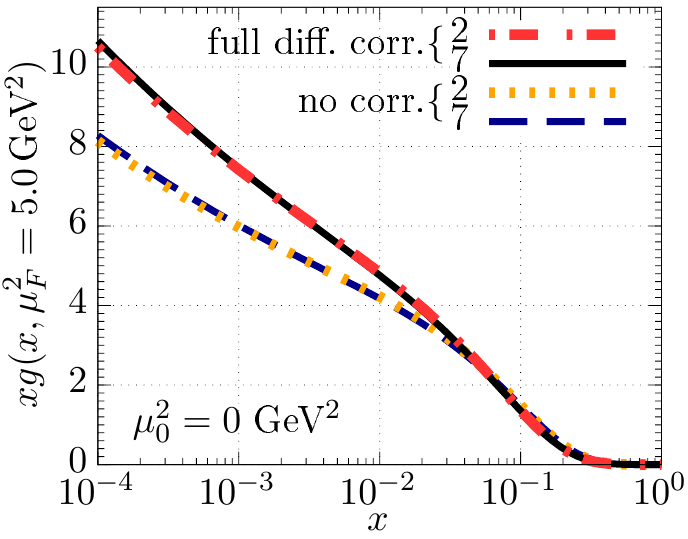}
		\vspace{-0.0cm}
		\caption{\sf The effect of absorptive corrections on the gluon density at scale $\mu_F^2=5$ GeV$^2$. At this relatively low scale, the gluons already grow with $x$ decreasing. The fits using data with $Q^2 >2$ GeV$^2$ and with $Q^2>7$ GeV$^2$ essentially gives the same gluon density, demonstrating the stability of the evolution from 2 to 7 GeV$^2$. }
		\label{fig:3}
	\end{center}
\end{figure}

\begin{figure} [tbh]
	\begin{center}
		\vspace*{0.0cm}
			\includegraphics[width=14cm]{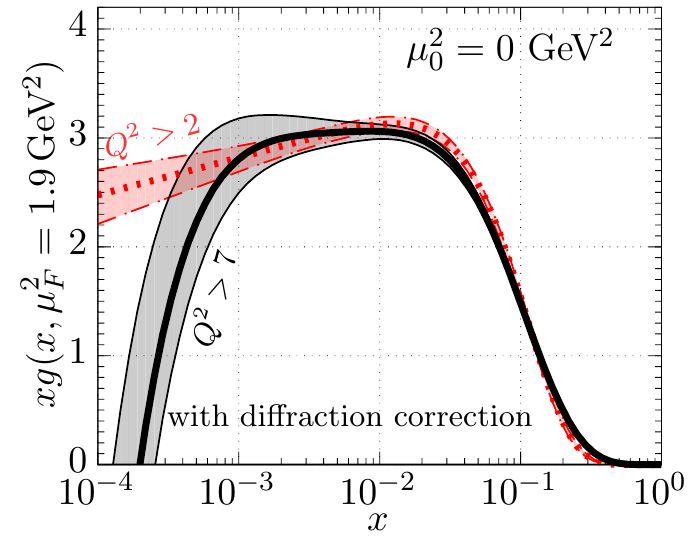}
		\vspace{-0.0cm}
		\caption{\sf Gluon distributions obtained from fits using (full) diffraction corrections and with a parameterization with the final (negative) term in (\ref{eq:p}). There is a large  instability for $x<10^{-3}$ depending on whether the DIS data in the region $2 < Q^2 < 7$ GeV$^2$ is included or not.}
		\label{fig:5}
	\end{center}
\end{figure}

First we consider the results obtained with $\mu_0=0$, that is for the case where the argument of the QCD coupling is not shifted, and the conventional NLO coupling normalized to $\alpha_s(M_Z^2)=0.118$ is used.

As we see from Fig.~\ref{fig:1}, accounting for absorptive corrections gives a noticeably larger `linear' gluon at low $x$.
We call it `linear' since after we correct the data points by adding the diffractive cross sections, in other words after we cancel (exclude from the data) the absorptive effect (at least its major part), we get the partons which should be described, indeed, by {\em linear} DGLAP evolution. These are not the partons which should be used to calculate completely inclusive cross sections based on the QCD factorization theorems. Thanks to AGK rules, in this inclusive case one has to take the conventional PDFs of the `global' fits in which the absorptive effects discussed above are included.

However these conventional partons should not be well described by linear DGLAP evolution; the non-linear corrections are essential in the low-$x$ and relatively low scale domain.
Thus here we restore the `original' partons from the proton wave function (that is partons not affected by power corrections). These partons should be better described by the linear DGLAP evolution and can be interpreted at $Q=Q_0$ as the original distributions in the incoming proton wave function.

Here we focus on the gluon, since at low-$x$ it is the most important PDF. At low-$x$ sea quarks distributions reproduce the same behaviour as that for gluons. The $x$ dependence of $xg$ is shown in Fig.~\ref{fig:1} at  a low scale $\mu^2=1.9$ GeV$^2$ which can be viewed as the input scale.  The plot contains three pairs of curves. The lower pair of curves are the conventional gluons, obtained ignoring absorptive corrections. 
They have a tendency to decrease with decreasing $x$. This may be caused by the fact that in fitting the data with a linear equation, the computer tries to mimic absorptive effects (which are not included in linear evolution) by the decrease of gluon densities in the input PDF. Excluding the negative absorptive correction we get larger `original' gluons. Each pair of curves was obtained by first fitting to those DIS data with $Q^2>2$ GeV$^2$, and then, secondly using the reduced data set with $Q^2>7$\,GeV$^2$. The fact that the results at starting scale 1.9 GeV$^2$ are almost the same reflects the stability of the fit and the fact that the evolution equation satisfactorily describes
the $Q^2$ dependence of the data within the 2 to 7 GeV$^2$ interval. 

The pair of upper (bold) curves in Fig.~\ref{fig:1}  show the gluon when the `{\it full}' absorptive effect was included, while the pair of thin curves are the result when only the `{\it elastic}' dDIS data are used (with the proton being left intact). The point is that the full absorptive correction includes the contribution of the proton excitations in the intermediate state (the cut of a lower line in Fig.~\ref{fig:A}). The probability of different $p\to N^*$ excitations is about 40\%. Therefore calculating the last term, $\sigma^{\rm dDIS}$, in (\ref{eq:correction}) we multiply the cross section given by the MRW fit by the factor 1.4. The increase in the gluon when going from a factor 1 to 1.4 is evident in Fig~\ref{fig:1}.

We emphasize that after power corrections arising from absorptive corrections are included the quality of the fit becomes appreciably better (see Table~\ref{tab:1}).
When we include the absorptive effect we gain $\Delta\chi^2/{\rm n.d.f.}=0.126$ (i.e. more than 148 in the total $\chi^2$ value; 1185 data points were fitted) in the fit with $Q^2>2$ GeV$^2$. 

For a moment let us digress to study the results obtained if we now include another source of power corrections. Namely, those coming
from the low scale behaviour of the running $\alpha_s$ coupling. If we perform a fit which includes a $\mu_0^2=1$\,GeV$^2$ shift of the argument of $\alpha_s$, then we obtain even larger `linear' gluons which become almost flat in the small $x$ region, especially when using the data at low $Q^2>2$ GeV$^2$,  close to the input; see Fig.~\ref{fig:2}. 
From Table~\ref{tab:1} you can see by comparing the $\chi^2$ values of the $\mu_0=1$ GeV fits with those of the $\mu_0=0$ fits that the inclusion of the `confinement' power corrections  practically does not change the quality of the fits. However they lead to a more acceptable flat low-$x$ behaviour of the `original' gluons\footnote{The low-$x$
behaviour of the gluon
is driven by the rightmost singularity of the amplitude
corresponding to 
vacuum quantum number exchange. In perturbative QCD the intercept of this
(QCD Pomeron)
 singularity is $\alpha_P(0)>1$. That is, we expect an increase of $xg(x)$ as $x$ decreases, accounting for absorptive effects. This leads to saturation of the gluon
density at low $x$ and
low $Q^2$, which may generate a flat behaviour of the input gluon. However there is no reason for
 $xg(x)$ to decrease as $x\to 0$.}. 

Returning to fits with $\mu_0=0$, we show, in Fig.~\ref{fig:3}, the gluons at a larger scale $\mu^2=5$ GeV$^2$. Already after a rather small interval of evolution (from 1.9 to 5 GeV$^2$) the gluons start to grow with $x$ decreasing. The results obtained by fitting data with relatively large $Q^2>7$ GeV$^2$ coincide well with those given by the fit to all the HERA data with $Q^2>2$ GeV$^2$. Again this stability demonstrates that the evolution equations well describe the $Q^2$ interval from 2 to 7 GeV$^2$.

In all the previous fits that we have shown, the input PDFs were parametrised by a form, (\ref{eq:p}) in which the gluons are definitely positive (that is, we set parameter $A'=0$). If we add the (parametric) term which allows for the  low $x$ gluons to be negative then we observe in Fig.~\ref{fig:5} a huge {\em instability} of the gluon density  for $x<10^{-3}$.
Here by
instability we mean that the gluon PDF strongly depends on the $Q^2$ interval
used to fit the data. Starting from $Q^2>7$ GeV$^2$ we obtain for $x<0.4\times
10^{-3}$ much smaller gluon densities than those which result from fitting data
in the larger $Q^2<2$ GeV$^2$ 
interval. As seen from Fig.~\ref{fig:5} the difference exceeds the error bars
formally coming from the fits.  Recall that without the term which allows
for negative gluons the results coming from the fits using the $Q^2< 2$ GeV$^2$ and
$Q^2<7$ GeV$^2$ intervals do not differ too much (see e.g. Fig.~\ref{fig:1}). 
Note also that when including the low $Q^2\sim 2$ GeV$^2$ HERA DIS data (which are close to the input) we get more or less the same results as in the previous (non-negative) fit (Fig.~\ref{fig:1}). On the other hand, however,  the fit using only the  $Q^2>7$ GeV$^2$ data results in a {\em negative} gluon for $x<2\times 10^{-4}$. In Table 1 this fit is denoted by `1.4~MRW~$-$'. We see that with two extra parameters ($A'$ and $B'$ in (\ref{eq:p})) the quality of the fits did not improve. Moreover, allowing for the gluon density to be negative we obtain completely unstable results at low $x$. 

Here we should emphasize that the error bands on the gluon distribution shown as examples in Figs. \ref{fig:2} and \ref{fig:5} have been obtained using the usual $\Delta\chi^2=1$ criteria. Due to the restrictive form of the input parametrization\footnote{Essentially $xg=Ax^B(1-x)^C$ with $A$ fixed by the momentum sum rule.} the error bands shown are not representative of the true uncertainty on the gluon distribution for $x\lapproxeq 10^{-3}$, which in this domain will be much larger if a more free or extensive parametrization were to be used. However, as far as the relative width of the error bands is concerned, Fig.~\ref{fig:5} does show the improvement in the uncertainty if all the HERA data with $Q^2 >2$ GeV$^2$ were fitted rather than the smaller subset with $Q^2 >7$ GeV$^2$.

\begin{table}
	\begin{center}
	\begin{tabular}{|c|c|c|c|c|c|c|c|}
		\hline $\mu_0$ (GeV) 
		& 0 & 0 & 1 & 1  \\ 
		\hline Data cut (GeV$^2$) 
		& 2.0 & 7.0 & 2.0 & 7.0  \\ 
		\hline
		\hline without~+ & \textbf{1.225} & 1.165 & 1.226 & 1.160  \\ 
		\hline without~$-$ & 1.227 & 1.161 & 1.223 & 1.157  \\ 
		\hline MRW~+    & 1.128 & 1.088 & 1.128 & 1.085  \\ 
		\hline 1.4~MRW~+ & 1.099 & 1.066 & \textbf{1.099 }& 1.063  \\ 
		\hline 1.4~MRW~$-$ & 1.099 & 1.065 & 1.093 & 1.063  \\ 
		\hline 
	\end{tabular} 
	\caption{The quality of the fits in terms of  reduced $\chi^2$ by degrees of freedom, that is $\chi^2/$n.d.f. MRW (1.4~MRW) denote that the elastic (full) diffractive correction has been made. `Without' means no diffractive correction has been made. The + ($-$) signs indicate positive (`negative') gluons; that is, in the parameterization of (\ref{eq:p}) the last term has been omitted (included). The largest improvement in the description of the DIS data  is obtained by the inclusion of (MRW) absorptive corrections. The favoured fits of the $Q^2>2$ GeV$^2$  HERA DIS data \cite{HERAdata} before and  after allowing for power corrections are shown in bold.}
	\label{tab:1}
\end{center}
	
\end{table}

\begin{table}\begin{center}
\begin{tabular}{|c|c|c|c|c|c|}
	
	\hline $\mu_0=0$ GeV      & $A$ & $B$ & $C$ & $D$ & $E$  \\ 
	\hline $x g$ & 5.63\phantom{0} * & 0.103 \phantom{*}& 10.261 & -- & --  \\ 
	\hline $x u_v$ & 3.743 *& 0.687 \phantom{*}& 4.822& -- & 14.087\\ 
	\hline $x d_v$    & 3.101 *& 0.799 \phantom{*}& 4.091& -- & --  \\ 
	\hline $x\bar U$    & 0.121 *& -0.169 *& 7.172 & 8.435 & --  \\
	\hline $\textstyle x \bar D$ & 0.201 \phantom{*}& -0.169 \phantom{*}& 5.732 & -- & -- \\ 
	\hline 
	
\end{tabular} 
\caption{\sf Parameters of the input parton distribution at the starting scale $\mu_F^2 = 1.9$\,Gev$^2$ using in the fit all data with $Q^2 > 2$\,GeV$^2$ and without changing the running of the coupling ($\mu_0 = 0$). Full diffractive corrections are included (1.4 MRW), there are no negative gluons (+) and the parameterization function is $A x^B (1-x)^C (1 + Dx + Ex^2)$. The parameters with an asterisk are not free since they are fixed by sum rules or the expected behaviour at low $x$.}
\label{tab:2}
\end{center}
\end{table}

We include in Tables~\ref{tab:2} and \ref{tab:3} the free parameters of the input parton distribution at the starting scale $\mu_F^2 = 1.9$\,GeV$^2$. Both sets of values are for the inclusion of the full diffractive corrections without negative gluons (1.4 MRW $+$). All data with $Q^2 > 2$\,GeV$^2$ are used and in Table~\ref{tab:2} the running of the coupling is not changed ($\mu_0 = 0$), while in Table~\ref{tab:3} it is ($\mu_0 = 1$\,GeV). The $A$ parameters for the gluon and valence quarks are fixed from momentum and quark number sum rules, respectively. All parameters in the down-type quark distribution $x \bar D$ are free. This distribution is assumed to be composed of 60\% down quark and 40\% strange quark. For the up-type distribution ($x\bar U = x\bar u$), $A$ and $B$ are fixed so that the small $x$ behavior of the up and the down quark distribution are the same, while $C_{\bar U}$ and $D_{\bar U}$ are kept free. No charm, bottom or top quarks are considered at the starting scale, and charm and bottom quark are generated perturbatively during the evolution. For more information, see Section 6.2 of Ref.~\cite{HERAdata}.

\begin{table}\begin{center}
\begin{tabular}{|c|c|c|c|c|c|}
	
	\hline $\mu_0=1$ GeV      & $A$ & $B$ & $C$ & $D$ & $E$   \\ 
	\hline $x g$ & 4.21\phantom{0} * & 0.021 \phantom{*}& 9.427 & -- & -- \\ 
	\hline $x u_v$ & 3.502 * & 0.665 \phantom{*}& 4.866 & -- & 15.066 \\ 
	\hline $x d_v$    & 2.939 * & 0.781 \phantom{*}& 4.054 & -- & -- \\ 
	\hline $x\bar U$  & 0.132 *  & -0.157 *& 6.936 & 7.078 & -- \\
	\hline $\textstyle x \bar D$ & 0.220 \phantom{*} & -0.157 \phantom{*} & 6.665 & --& -- \\ 
	\hline 
\end{tabular} 
\caption{\sf Free parameters of the input parton distribution at the starting scale $\mu_F^2 = 1.9$\,Gev$^2$ using in the fit all data with $Q^2 > 2$\,GeV$^2$ and running of the coupling is changed ($\mu_0 = 1$\,GeV$^2$).  Full diffractive corrections are included (1.4 MRW), there are no negative gluons (+) and the parameterization function is $A x^B (1-x)^C (1 + Dx + Ex^2)$. The smaller value of $B$ (as compared with $B=0.103$ in Table~\ref{tab:2}) makes the gluons more flat. The parameters with an asterisk are not free since they are fixed by sum rules or the expected behaviour at low $x$.}
\label{tab:3}
\end{center}
\end{table}

\section{Summary}
We have investigated the effects of absorption, which has the form of a power correction, at the beginning of low $Q^2$ evolution. Using the AGK cutting rules and the known diffractive part of the DIS cross section, we are able to remove the absorptive corrections from the inclusive data. Interestingly this greatly  {\it improves}  the description of the data. The gluons corresponding to this linear DGLAP evolution are significantly {\it larger} at small $x$ and small $Q^2$.  Moreover they are {\it stable} to the low $Q^2$ cut used to select the DIS data to be fitted~\footnote{The role of higher twist=4 contributions in the description
of low $x$ and low $Q^2$ PDFs was studied in \cite{MOT}. Gluon PDFs larger than those found in conventional global analyses were obtained in this
domain. However, while we have used the known diffractive DIS data to
account for absorptive effects {\em during} the evolution, the author of
\cite{MOT} put a theoretically motivated ansatz for the twist=4
contribution at the final $Q^2$ corresponding  to the PDF scale. The difference is that we
 actually study not twist=4 but twist=2 parton distributions
    accounting for the (induced by twist=4) absorptive corrections which
    modify the twist=2 evolution through the addition of a known
   inhomogeneous term.}.

A second power correction arises from the freezing of $\alpha_s$ at low $Q^2$ as was proposed in \cite{Simonov:2001dt,Shir}. The main effect is to make the input gluon distribution essentially flat at very low $x$.

Recall that the conventional `global' PDFs should be used to describe inclusive cross sections, based on the QCD factorization theorems. However, to study theoretically the details of the proton wave function, necessary for the description of exclusive data, one has to use these new gluon densities based on linear evolution, not affected by absorptive and other power corrections.

Recent studies \cite{BFKL1,BFKL2} have added
to the DGLAP splitting
functions the contribution given by the low-$x$ re-summation of BFKL-like
($(\alpha_s\ln(1/x))^n $)
terms. It was found that this improves the conventional (linear) DGLAP fit of the low
$x$ and $Q^2$ data.
 This observation does not mean that non-linear effects should be neglected.
Both the power corrections caused by the inhomogeneous (non-linear) terms in DGLAP
evolution (\ref{eq:1}), ({\ref{a1}), and also the re-summation of the higher $\alpha_s$ order leading
logarithmic (in $\ln(1/x)$)
contributions are important in the low $x$ and $Q^2$ domain. It would be
interesting in future to
 fit the data accounting for both effects. However this is beyond the scope of the present paper.

\section*{Appendix}
Here we briefly recall the main elements/features of the MRW~\cite{MRW,MRW1} description of diffractive parton distributions (that is dDIS PDFs). In the MRW approach the inhomogeneous term arises from the perturbative Pomeron initiated contribution, that is from the fusion of two parton cascades into the one cascade. Then
the evolution of a diffractive PDF, $a^D$, with scale $\mu^2$, takes the form
\begin{equation}
	\label{a1}
	\frac{da^D(x_P,z,\mu^2)}{d{\rm ln}\mu^2}=\sum_{a'=q,g} P_{aa'}\otimes a'^D+\sum_{P=G,S,GS}P_{aP}(z)f_P(x_P;\mu^2)\ .
\end{equation}
The Pomeron cascade  contains gluon and quark components which are described by the conventional gluon, $xg(x,\mu^2)$, and sea quark, $xS(x,\mu^2)$, PDFs. Besides this splitting of the gluon-made and the quark-made pomerons is possible. These contributions are denoted as $G,S$ and $GS$ respectively. The functions $f_{P=G,S,GS}(x_P,\mu^2)$ in (\ref{a1}) are the fluxes of the respective Pomerons, while  $P_{aP}(z)$ describes the parton $a$ distribution (over the Pomeron momentum fraction $z$) corresponding to the splitting of the $P=G,S,GS$  component of the Pomeron to the parton $a$;
$P_{aa'}$ are the usual splitting functions of parton $a'$ to parton $a=q,g$.
The values of fluxes are given by
\be
\label{G}
f_{P=G}(x_P,\mu^2) = \frac 1{x_PB_D}\left[R_g\frac{\alpha_s(\mu^2)}{\mu}x_Pg(x_P,\mu^2)\right]^2\ ,
\ee
\be 
\label{S}
f_{P=S}(x_P,\mu^2) = \frac 1{x_PB_D}\left[R_q\frac{\alpha_s(\mu^2)}{\mu}x_Pq(x_P,\mu^2)\right]^2\ ,
\ee
\be
\label{GS}
f_{P=GS}(x_P,\mu^2) = \frac 1{x_PB_D} R_gR_q\frac{\alpha^2_s(\mu^2)}{\mu^2}x_Pg(x_P,\mu^2)x_Pq(x_P,\mu^2)\ .
\ee
Here $x_P$ is the proton momentum fraction carried by the Pomeron, factor $R_{g,q}$ accounts for the skewness, that is for the fact that actually the Pomeron is described by the generalized distribution function (GPD) with the non-zero momentum transferred through the Pomeron, where the fraction $x'$ carried on one side of the ladder can  differs from that, $x=x_P$ on the other side. In our case we deal with $x'\ll x_P$.
Finally  the diffractive slope $B_D$ plays the role of $R^2$ in (\ref{eq:1}).

The functions $P_{aP}(z)$, which plays the role of `initial' parton distributions produced by the Pomeron at scale $\mu$, were calculated at lowest $\alpha_s$ order in the Appendix of ~\cite{MRW1}.
They read:
\be
\label{qG}
P_{qG}(z)=z^3(1-z)\ ,
\ee 
\be\label{gG}
P_{gG}(z)=\frac 9{16}(1-z)^2(1+2z)^2\ ,
\ee
\be
\label{qS}
P_{qS}(z)=\frac 4{81}z(1-z)\ ,
\ee 
\be\label{gS}
P_{gS}(z)=\frac 19(1-z)^2\ ,
\ee
\be
\label{qGS}
P_{qGS}(z)=\frac 29z^2(1-z)\ ,
\ee 
\be\label{gGS}
P_{gGS}(z)=\frac 14(1-z)^2(1+2z)\ .
\ee
These expressions are used during the evolution when  $\mu>Q_0$.  The part coming from scales $\mu<Q_0$ lower than the input value $Q_0$, was parametrized as the usual input PDF.

It is seen from (\ref{G},\ref{S},\ref{GS}) that the contribution of the inhomogeneous term in (\ref{a1}) is suppressed by the factors $\alpha_s^2$ and $1/B_D\mu^2$. That is it should be considered as a power correction  ($1/\mu^2$), and so dies out at large $\mu$. However at low $x$ it is strongly enhanced by the large value of parton densities, like  $(xg)^2$. Therefore its contribution is not negligible and it affects the beginning of the conventional linear DGLAP evolution.

\section*{Acknowledgments}

We thank Lucian Harland-Lang, Robert Thorne and Graeme Watt for discussions on this issue. MGR and EGdO thank the IPPP at the University of Durham for hospitality. This work was supported by Fapesc, INCT-FNA (464898/2014-5), and CNPq (Brazil) for EGdO and MRP. This study was financed in part by the Coordena\c{c}\~ao de Aperfei\c{c}oamento de Pessoal de N\'ivel Superior -- Brasil (CAPES) -- Finance Code 001.

\end{document}